\documentclass[aps,prb,twocolumn,amsmath,amssymb,superscriptaddress]{revtex4-2}

\usepackage{siunitx}
\usepackage[version=4]{mhchem}
\usepackage{graphicx}

\usepackage{bm} 
\renewcommand*{\vec}[1]{\bm{#1}}

\usepackage[version=2]{acro}
\acsetup{single=false}

\DeclareAcronym{fm}{short = FM, long = ferromagnet, short-indefinite ={an}, long-indefinite ={a}}
\DeclareAcronym{afm}{short = AFM, long = antiferromagnet, short-indefinite ={an}, long-indefinite ={an}}

\DeclareAcronym{dmi}{short = DMI, long = Dzyaloshinskii--Moriya interaction}
\DeclareAcronym{soc}{short = SOC, long = spin-orbit coupling}

\DeclareAcronym{nsot}{short = NSOT, long = N\'{e}el-spin-orbit torques, short-indefinite ={an}, long-indefinite ={a}}

\DeclareAcronym{llg}{short = LLG, long = Landau--Lifshitz--Gilbert, short-indefinite ={an}, long-indefinite ={a}}
\DeclareAcronym{skkr}{short = SKKR, long = screened Korringa--Kohn--Rostoker, short-indefinite ={an}, long-indefinite ={a}}
\DeclareAcronym{scf}{short = SCF, long = self-consistent field, short-indefinite ={an}, long-indefinite ={a}}
\DeclareAcronym{rsp}{short = RSP, long = relativistic spin-polarized, short-indefinite ={an}, long-indefinite ={a}}
\DeclareAcronym{rtm}{short = RTM, long = Relativistic Torque Method, short-indefinite ={an}, long-indefinite ={a}}

\DeclareAcronym{asa}{short = ASA, long = atomic sphere approximation}

\DeclareAcronym{dlm}{short = DLM, long = Disordered Local Moment}
\DeclareAcronym{rdlm}{short = RDLM, long = relativistic Disordered Local Moment}
\DeclareAcronym{sce}{short = SCE, long = Spin-Cluster Expansion}

\DeclareAcronym{dos}{short = DOS, long = density of states}

\begin{document}

\title{Current induced switching in \ce{Mn2Au} from first principles}

\author{Severin Selzer}
\affiliation{Fachbereich Physik, Universität Konstanz, DE-78457 Konstanz, Germany}

\author{Leandro Salemi}
\affiliation{Department of Physics and Astronomy, Uppsala University, P.\,O.\ Box 516, S-751 20 Uppsala, Sweden}

\author{Andr\'{a}s De\'{a}k}
\affiliation{Department of Theoretical Physics, Institute of Physics, Budapest University of Technology and Economics, M\H uegyetem rkp.\ 3.,	HU-1111 Budapest, Hungary}

\author{Eszter Simon}
\affiliation{Department of Theoretical Physics, Institute of Physics, Budapest University of Technology and Economics, M\H uegyetem rkp.\ 3.,	HU-1111 Budapest, Hungary}

\author{L\'{a}szl\'{o} Szunyogh}
\affiliation{Department of Theoretical Physics, Institute of Physics, Budapest University of Technology and Economics, M\H uegyetem rkp.\ 3.,	HU-1111 Budapest, Hungary}
\affiliation{MTA-BME Condensed Matter Research Group, Budapest University of Technology and Economics, M\H uegyetem rkp.\ 3., HU-1111 Budapest, Hungary}

\author{Peter M. Oppeneer}
\affiliation{Department of Physics and Astronomy, Uppsala University, P.\,O.\ Box 516, S-751 20 Uppsala, Sweden}

\author{Ulrich Nowak}
\affiliation{Fachbereich Physik, Universität Konstanz, DE-78457 Konstanz, Germany}

\date{\today}

\begin{abstract}
It is well established that it is possible to switch certain antiferromagnets electrically, yet the interplay of \ac{nsot} and thermal activation is only poorly understood. Combining \emph{ab initio} calculations and atomistic spin dynamics simulations we develop a multiscale model to study the current induced switching in \ce{Mn2Au}. We compute from first principles the strength and direction of the electrically induced magnetic moments, caused by the Rashba--Edelstein effect, and take these into account in atomistic spin dynamics simulations. Our simulations reveal the switching paths as well as the time scales for switching. The size of the induced moments, however, turns out to be insufficient to lead to fully deterministic switching. Instead, we find that a certain degree of thermal activation is required to help overcoming the relevant energy barrier. 
\end{abstract}


\maketitle

\section{Introduction}

\Acp{afm} are promising materials for spintronic devices. Among the advantages over \acp{fm} are the lack of stray fields, the very low susceptibility to magnetic fields, the abundance of materials and much faster spin dynamics \cite{jungwirth_antiferromagnetic_2016,zelezny_spin_2018,baltz_antiferromagnetic_2018}. However, the antiferromagnetic order parameter in \acp{afm} is difficult to read and to control because of a lack of macroscopic magnetization, a fact which is strongly related to some of their advantages. A major step in the field of antiferromagnetic spintronics \cite{jungwirth_antiferromagnetic_2016,zelezny_spin_2018,baltz_antiferromagnetic_2018}
was the discovery of electrically induced \ac{nsot} \cite{zelezny_relativistic_2014,wadley_electrical_2016,olejnik_antiferromagnetic_2017,bodnar_writing_2018,olejnik_terahertz_2018} in specific antiferromagnetic materials. 
These torques are a result of a special magnetic structure, where, for the magnetic state, global inversion symmetry is broken but one sublattice forms the inversion partner of the other, in combination with the inverse spin-galvanic or (Rashba--)Edelstein effect \cite{zelezny_relativistic_2014}, which is the generation of a nonequilibrium spin polarization by electrical currents.
Currently, \ce{CuMnAs} and \ce{Mn2Au} are the two known materials that provide antiferromagnetic order at room temperature and possess the specific crystal structure required for \ac{nsot}. The latter is the more promising material as its critical temperature is extremely high---higher than the peritectic temperature of about \SI{950}{\kelvin}, where the material decomposes \cite{barthem_revealing_2013}---and it is easier to handle due to the lack of toxic components.

Despite the fact that several studies clearly demonstrate that it is possible to switch the order parameter of \ce{Mn2Au} via the application of an electrical current by $90^{\circ}$ \cite{zelezny_relativistic_2014,roy_robust_2016,meinert_electrical_2018,bodnar_writing_2018,salemi_orbitally_2019}, the switching mechanism---whether deterministic or thermally activated, coherent or via domain wall motion---remains concealed. The employed models and simulations so far rest on phenomenological descriptions \cite{roy_robust_2016} and macrospin approximations \cite{meinert_electrical_2018}. A microscopic and quantitative model of the switching process is missing. 

Here, we combine \emph{ab initio} calculations with atomistic spin dynamics simulations to develop and employ a multi-scale model of the current induced switching in \ce{Mn2Au}.
The three ingredients for this multi-scale model are \emph{ab initio} calculations of the exchange interactions and anisotropies (section II), first-principles calculations of the current induced magnetic moments (section III), and atomistic spin model simulations (section IV), that include the results from the first-principles calculations and investigate the switching mechanism and its dynamics. We show that the switching is fast, on a time scale of some tens of picoseconds, but not purely deterministic, requiring some degree of thermal activation to overcome the anisotropy energy barrier during the switching process.

\section{Derivation of the spin model from ab initio calculations} 
\begin{figure}
	\centering
	\includegraphics[width=\linewidth]{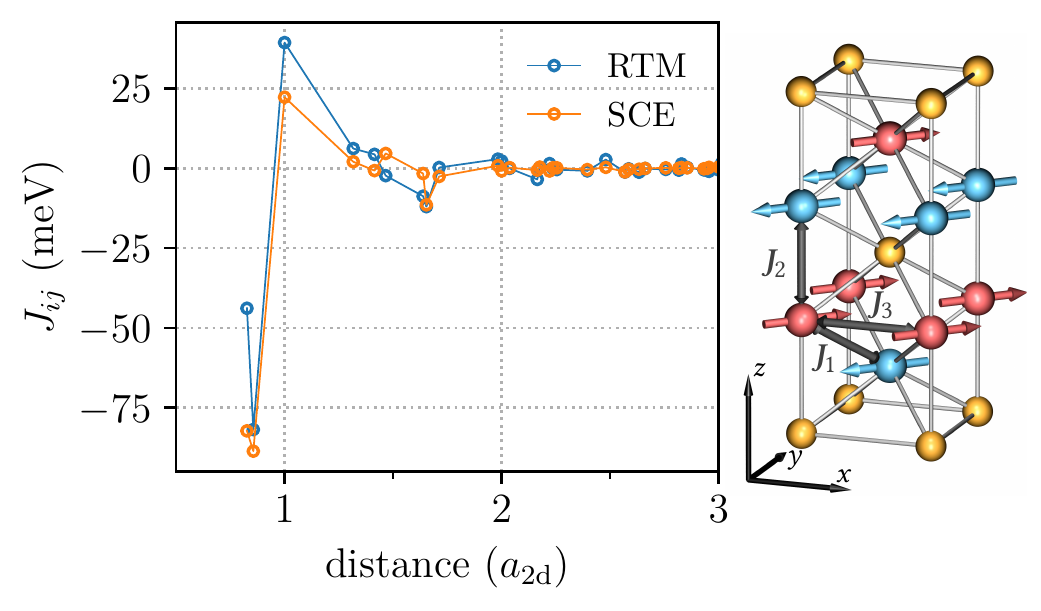}
	\caption{Left: Isotropic exchange interactions as a function of distance calculated by using the \acs*{rtm} and the \acs*{sce} methods. Right: Crystal structure of \ce{Mn2Au}. The two \ce{Mn} sublattices are illustrated by red and blue spheres. The ground state orientations of the magnetic moments are indicated by arrows. The first three nearest neighbor magnetic exchange interactions $J_i$ are also visualized in the figure. \label{fig:exchange}}
\end{figure}

We employ the fully relativistic \ac{skkr} method  \cite{zabloudil_electron_2005} to determine the electronic structure and magnetic interactions of \ce{Mn2Au}. \ce{Mn2Au} crystallizes in the \ce{MoSi2} structure with the lattice constants $a_{\mathrm{2d}}=\SI{3.328}{\angstrom}$ and $c=\SI{8.539}{\angstrom}$ \cite{wells_structure_1970,shick_spin-orbit_2010,barthem_revealing_2013}.
The \ce{MoSi2}-type lattice geometry is depicted in Fig.~\ref{fig:exchange}. The potentials were treated within the \ac{asa} with an angular momentum cutoff of $\ell_\text{max}=2$ to describe the electron scattering. For energy integrations we used 15 energy points on a semicircular contour on the upper complex semiplane, and up to 7260 
$k$-points in the irreducible wedge of the Brillouin zone near the Fermi energy for the calculation of spin model parameters.

We perform self-consistent calculations for the layered \ac{afm} state shown in Fig.~\ref{fig:exchange}, which has been identified as the magnetic ground state by neutron diffraction experiments \cite{barthem_revealing_2013},
but also for the \ac{fm} state.
We find the layered \ac{afm} state lower in energy than the \ac{fm} state by 25.8 mRy/atom, which compares fairly well to the value reported in Ref.~\cite{khmelevskyi_layered_2008} (21.5 mRy/atom).
Also in agreement with Ref.\ \cite{khmelevskyi_layered_2008} we obtain a larger magnetic moment for the \ce{Mn} atoms in the layered \ac{afm} state ($\SI{3.74}{\mu_{\mathrm{B}}}$) than in the \ac{fm} state ($\SI{3.70}{\mu_{\mathrm{B}}}$).
For the description of the switching process we consider the following spin model:
\begin{equation}
	\begin{aligned}
		\mathcal{H} =&- \frac{1}{2}\sum_{i\neq j}J_{ij}\vec{S}_{i}\cdot\vec{S}_{j}
		-\sum_{i} d_{z}S_{i,z}^2\\
		&-\sum_{i} d_{zz}S_{i,z}^4
		-\sum_{i} d_{xy}S_{i,x}^2S_{i,y}^2 \ ,
	\end{aligned}
        \label{eq:spin_model}
\end{equation}
where the isotropic exchange interactions $J_{ij}$ are obtained from 
the \ac{rtm} \cite{udvardi_first-principles_2003},
while the anisotropy parameters $d_{z}$, $d_{zz}$ and $d_{xy}$ are derived from band energy calculations in the spirit of the magnetic force theorem \cite{weinberger_magnetic_2009}. 

The isotropic exchange interactions calculated from the layered \ac{afm} state as reference are plotted in Fig.~\ref{fig:exchange} as a function of the interatomic distance. We can identify three dominant Heisenberg couplings: antiferromagnetic ones for the two nearest neighbors, $J_{1}=\SI{-43.84}{\milli\electronvolt}$ and $J_{2}=\SI{-81.79}{\milli\electronvolt}$, but a ferromagnetic one for the third nearest neighbor, $J_{3}=\SI{39.28}{\milli\electronvolt}$. 
These values show good qualitative agreement with those calculated in Ref.~\cite{khmelevskyi_layered_2008} also in terms of the KKR-ASA method, but using a cutoff of $\ell_{\rm max}=3$ for the partial waves,
$J_{1}=\SI{-68.30}{\milli\electronvolt}$,
$J_{2}=\SI{-91.70}{\milli\electronvolt}$ and
$J_{3}=\SI{19.86}{\milli\electronvolt}$.
Since the interactions $J_1$ and $J_2$ act between sublattices (layers), while $J_3$ is the leading interaction within a sublattice (cf.\ Fig.~\ref{fig:exchange}), these couplings clearly favor the layered \ac{afm} state as the ground state of the system.

It turns out that taking into account only the first three nearest neighbor interactions is not sufficient for a precise determination of the inter- and intra-sublattice interactions. In our simulations we, hence,  consider interactions up to a distance of $2.7\ a_{\mathrm{2d}}$, resulting in an inter-sublattice exchange interaction of $J_{\mathrm{inter}}=\SI{-371.13}{\milli\electronvolt}$ and an intra-sublattice exchange interaction of
$J_{\mathrm{intra}}=\SI{182.36}{\milli\electronvolt}$. 
Considering exchange interactions only in the first three shells yields $J_{\mathrm{inter}}=4J_1+J_2=\SI{-257.15}{\milli\electronvolt}$ and $J_{\mathrm{intra}}=4J_3=\SI{157.12}{\milli\electronvolt}$, being thus 30 and 14 \% smaller in magnitude than the ones calculated with a spatial cutoff of $2.7\ a_{\mathrm{2d}}$.

Experimental values for the effective inter-sublattice exchange coupling, $J_{\mathrm{eff}}=-J_{\mathrm{inter}}/4$ \cite{sapozhnik_experimental_2018},  were previously provided based on susceptibility measurements for \ce{Mn2Au} powder \cite{barthem_revealing_2013} and thin films \cite{sapozhnik_experimental_2018}, $J_{\mathrm{eff}}=\SI{75}{\milli\electronvolt}$ and $J_{\mathrm{eff}}=\SI{22 \pm 5}{\milli\electronvolt}$, respectively. The corresponding values from our calculations, $J_{\mathrm{eff}}=\SI{92.8}{\milli\electronvolt}$, and the one derived from the exchange interactions in Ref.~\onlinecite{khmelevskyi_layered_2008}, $J_{\mathrm{eff}}=\SI{90}{\milli\electronvolt}$, compare remarkably well and are also in good agreement with the experimental result for the powder sample \cite{barthem_revealing_2013}.    

From our spin dynamics simulations we obtain a N\'{e}el temperature of $\SI{1680\pm3}{\kelvin}$, which is in good agreement with the value of \SI{1610\pm10}{\kelvin} calculated in Ref.~\cite{khmelevskyi_layered_2008} via Monte-Carlo simulations using nine nearest neighbor shells (the numerical values of which, however, were not provided beyond the first three shells).
Note that due to a peritectic temperature of \SI{950}{\kelvin}, the N\'eel temperature can only be extrapolated from experiments, yielding values in the range of \SIrange{1300}{1600}{\kelvin} \cite{barthem_revealing_2013}.

In order to support the validity of our spin model description relying on the assumption of rigid magnetic moments that are stable against magnetic disorder, we also perform calculations using the \ac{rdlm} theory \cite{gyorffy_dlm_1985,staunton_rdlm_2006}. This approach assumes a fully spin disordered reference state, and also enables the extraction of spin model parameters by means of the so-called \ac{sce} \cite{drautz_spin-cluster_2004,szunyogh_atomistic_2011}, which maps the adiabatic magnetic energy surface onto a spin model.

The resulting isotropic Heisenberg couplings are also displayed in Fig.~\ref{fig:exchange}. There is a remarkable similarity between the two spin model parameter sets, despite their quantitative differences especially for the first and third neighbor shells. 
Obviously, the interactions obtained from the \ac{sce}-\ac{rdlm} calculation are also consistent with the layered \ac{afm} structure as ground state and we obtain a N\'{e}el temperature of $\SI{1786\pm3}{\kelvin}$, which is in good agreement with the \ac{rtm}. 

\begin{figure}
	\centering
	\includegraphics[width=\linewidth]{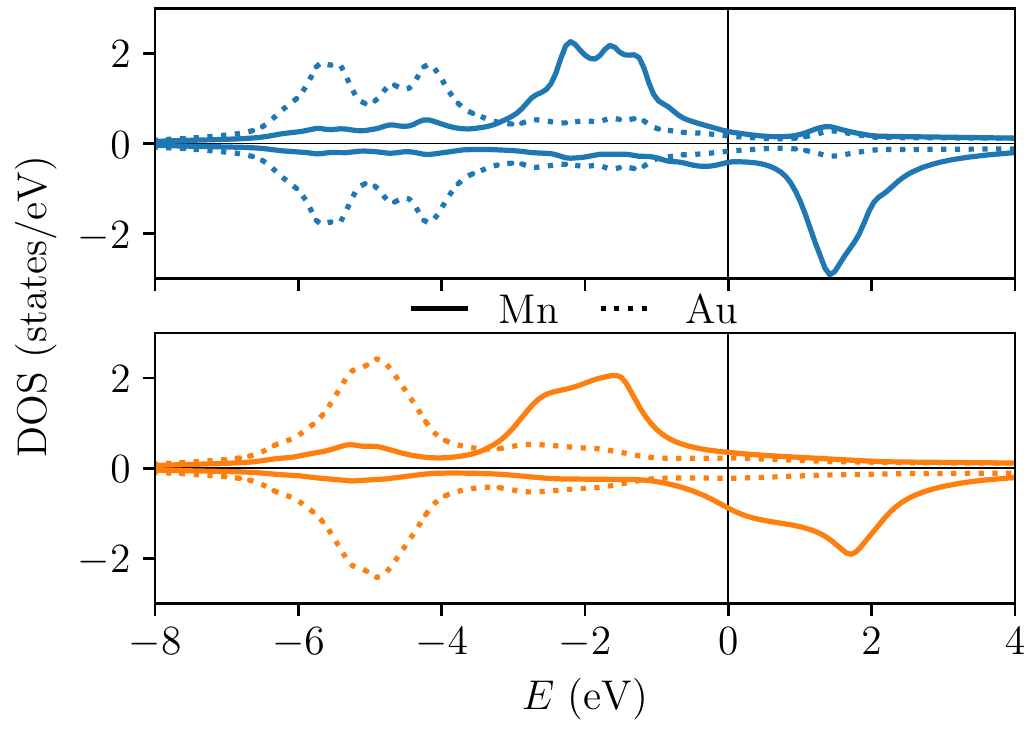}
	\caption{Density of states per atom from the electronic structure calculations in the \ac{afm} (top panel) and \ac{dlm} (bottom panel) states. The DOS for only one \ce{Mn} sublattice is shown. Positive values correspond to spin up states, negative ones to spin down states. \label{fig:DOS}}
\end{figure}

Conceptually, the \ac{rtm} gives a good approximation near the ground state, whereas the \ac{sce} corresponds to a high-temperature phase. The fact that the two sets of parameters agree well despite this fundamental difference between the two methods can be explained by the rigidity of the \ce{Mn} local spin moments.
In order to support this point we compare the \ac{dos} for the two magnetic states in Fig.~\ref{fig:DOS}. As also noted in Ref.~\cite{khmelevskyi_layered_2008}, the narrow bandwidth of the \ce{Mn} $\mathrm{d}$-bands and the formation of a pseudogap around the Fermi level are visible in the \ac{afm} state. The expected smearing of the \ac{dos} in the \ac{dlm} state due to spin disorder is clearly seen in the bottom panel of  Fig.~\ref{fig:DOS}, but the large exchange splitting between the two spin channels prevails. 
This shows up also in the spin moment of \ce{Mn} calculated in the \ac{dlm} state of $\SI{3.71}{\mu_{\mathrm{B}}}$ being practically the same as in the layered AFM state.

As for the anisotropies in Eq.~\eqref{eq:spin_model}, we calculate a second order anisotropy of $d_{z}=\SI{-0.62}{\milli\electronvolt}$, and fourth order anisotropies $d_{zz}=\SI{-0.024}{\milli\electronvolt}$ and $d_{xy}=\SI{0.058}{\milli\electronvolt}$. These values compare fairly well to those that can be derived from the anisotropy constants reported in Ref.~\cite{shick_spin-orbit_2010},  
$d_{z}=\SI{-1.19}{\milli\electronvolt}$, $d_{zz}=\SI{-0.015}{\milli\electronvolt}$, and $d_{xy}=\SI{0.04}{\milli\electronvolt}$, in particular considering that the latter ones were calculated in terms of a full-potential density functional method  contrary to the \ac{asa} we used in our calculations. This result is also in agreement with experimental reports of an upper bound for the in-plane anisotropy $d_{xy}$ of $\SI{0.068}{\milli\electronvolt}$ \cite{sapozhnik_direct_2018}.
Thus, in agreement with Refs.~\cite{shick_spin-orbit_2010,barthem_easy_2016,sapozhnik_direct_2018} we find the magnetic easy axis along the $\langle110\rangle$ direction as illustrated in Fig.~\ref{fig:exchange}. However, in our results the anisotropy responsible for the confinement in the basal plane is only about half in magnitude as compared to Ref.~\cite{shick_spin-orbit_2010}. Note though that the out-of-plane anisotropy plays only a minor role in the switching process discussed in our work. 

For our atomistic spin dynamics simulations we combine these anisotropies with the \ac{rtm} exchange parameters since both are calculated from the same converged potential in contrast to the \ac{sce} exchange parameters.

\section{First-principles calculations of the induced moments}
The inverse spin-galvanic or Rashba--Edelstein effect leads to electrically induced magnetic moments. These induced spin and orbital polarizations can be computed using the Kubo linear-response formalism.
Specifically, the locally induced polarizations can be expressed as
\begin{equation}
    \delta \vec{S} = \vec{\chi}^S \vec{E}, ~{\textrm{and}} ~~ \delta \vec{L} = \vec{\chi}^L \vec{E},
\end{equation}
with $\boldsymbol{\chi}^{S}$ and $\boldsymbol{\chi}^{L}$ 
the spin and orbital Rashba--Edelstein  susceptibility tensors, respectively,
and $\vec{E}$ the applied electric field.
The magneto-electric susceptibility tensors can be obtained by evaluating the response to a perturbing electric field, $\hat{V} = -e \hat{\vec{r}} \cdot \vec{E}$ where $e$ is the electron charge.

Employing DFT-based single-electron states, the susceptibility tensors are given by \cite{salemi2021}
\begin{align}
\label{eq:LinearResponse}
\chi^{S,L}_{ij} &= -\frac{ie}{m_e} \int_{\Omega} \frac{d\vec{k}}{\Omega}
\sum_{n\neq m} \frac{f_{n\vec{k}} - f_{m\vec{k}} }{\hbar \omega_{nm\vec{k}}}~
\frac{A^{(S,L)i}_{mn\vec{k}} ~ p^j_{nm\vec{k}} }{-\omega_{nm\vec{k}} + i\tau_{\text{inter}}^{-1}} \nonumber \\
& ~~~ -\frac{ie}{m_e} \int_{\Omega} \frac{d\vec{k}}{\Omega}
\sum_{n} \frac{\partial f_{n\vec{k}} }{\partial \epsilon}~
\frac{A^{(S,L)i}_{nn\vec{k}} ~ p^j_{nn\vec{k}} }{i\tau_{\text{intra}}^{-1}} \, .
\end{align}
Here,  $ \hbar \omega_{nm\vec{k}} = \epsilon_{n\vec{k}} - \epsilon_{m\vec{k}}$, with $\epsilon_{n\vec{k}}$ the unperturbed relativistic Kohn--Sham single electron energies, $\Omega$ is the Brillouin zone volume, $p^j_{nm\mathbf{k}}$ is the matrix element of the $j^\text{th}$ component of the momentum-operator, and
$f_{n\vec{k}}$ is the occupation of Kohn--Sham state $|n\vec{k}\rangle$. $\vec{A}^{(S,L)}_{mn\vec{k}}$ stands for a matrix element of the spin or orbital angular momentum operator, i.e., 
$\vec{A}^{S}_{mn\vec{k}} = \hat{\vec{S}}_{mn\vec{k}}$  for $\vec{\chi}^S$ and $\vec{A}^{L}_{mn\vec{k}} = \hat{\vec{L}}_{mn\vec{k}}$ for $\vec{\chi}^L$.
The parameters $\tau_{\text{inter}}$ and $\tau_{\text{intra}}$ are the electronic lifetimes for inter- and intraband scattering processes, respectively. These parameters capture the decay of an electron state $|n\vec{k}\rangle$ due to electron-electron scattering and interactions with external baths, e.g., phonons and defect scattering. In this work, we use an effective decay $\tau = \tau_{\text{inter}} = \tau_{\text{intra}} = \SI{50}{\femto\second}$.

To compute the current induced spin and orbital polarizations on the individual atoms in \ce{Mn2Au} we employ the relativistic DFT package WIEN2k \cite{Blaha2018}, which gives the Kohn--Sham energies $\epsilon_{n\vec{k}}$ and wave functions $|n\vec{k}\rangle$ that are then used in Eq.\ (\ref{eq:LinearResponse}). We calculate the induced magnetic moments 
for different orientations of the electrical field with respect to the magnetic easy axes, as reversible switching was reported for both the \mbox{[110]} and \mbox{[100]} directions \cite{bodnar_writing_2018}. Furthermore, we evaluate both the induced spin and orbital polarizations. 
The local magnetic moments induced by the electric field are finally given as
$\vec{\mu}= \vec{\mu}_S + \vec{\mu}_L = (2\vec{\chi}^S + \vec{\chi}^L)\vec{E}$.

\begin{figure*}[ht]
	\begin{minipage}[H]{.48\linewidth}
		\includegraphics[width=0.9\linewidth]{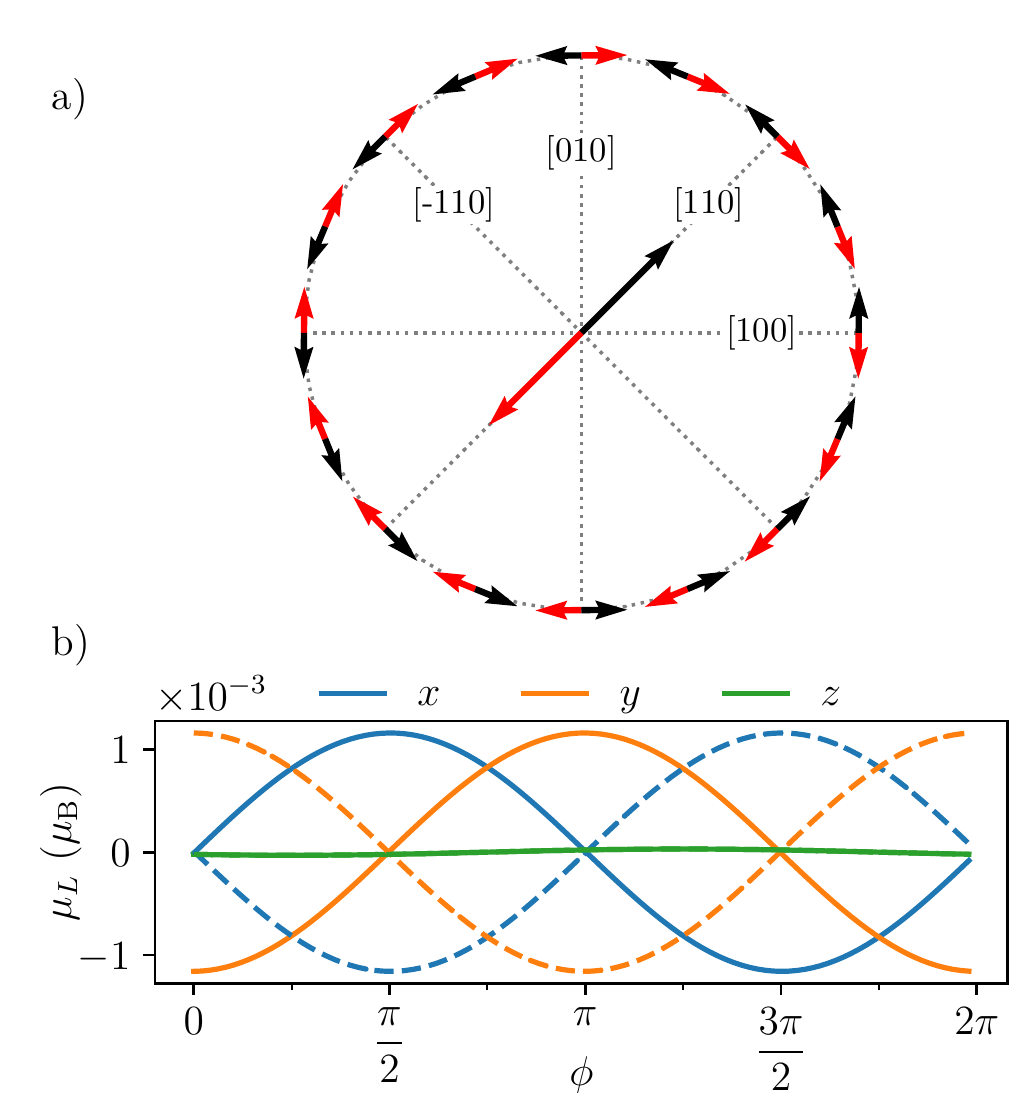}
	\end{minipage}%
	\begin{minipage}[H]{.03\linewidth}
		\hspace{\linewidth}
	\end{minipage}%
	\begin{minipage}[H]{.48\linewidth}
		\includegraphics[width=0.9\linewidth]{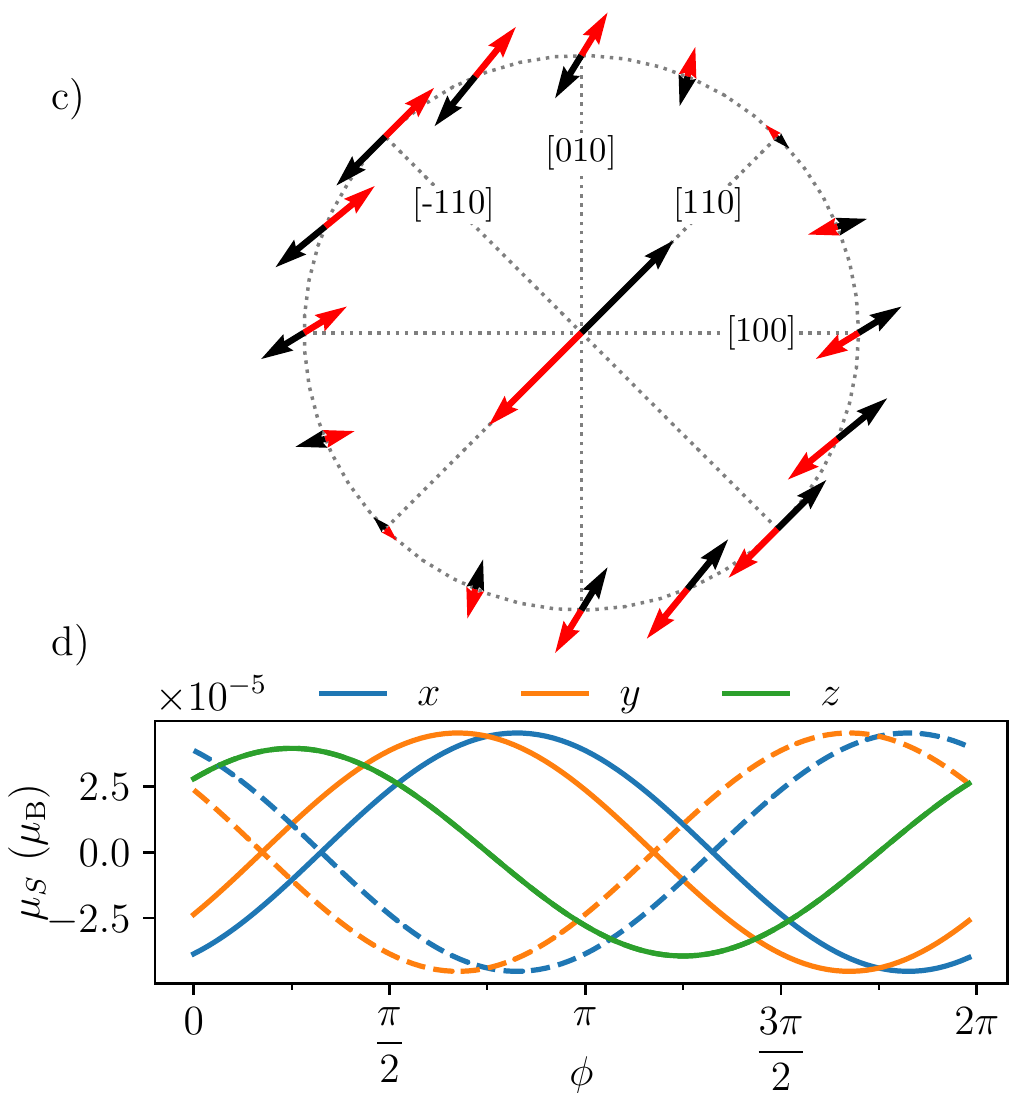}
	\end{minipage}
	\caption{Calculated induced orbital ($\mu_L$) and spin ($\mu_S$) moments on the two \ce{Mn} sublattices as a function of the electric field direction for a field of $E=\SI{1e7}{\volt\per\meter}$ and local magnetic moments oriented along the \mbox{[110]} direction.
	a) In-plane direction of the induced orbital moments on the two \ce{Mn} sublattices (black vs.\ red). The arrows in the center depict the local magnetic moments. b) Cartesian components of the induced orbital moments on the two \ce{Mn} sublattices (solid vs.\ dashed) as a function of the in-plane angle of the electric field with respect to the \mbox{[100]} axis. c) and d) same as in a) and b) for the induced spin moments. \label{fig:induced_moments}}
\end{figure*}

The calculated induced orbital and spin magnetic moments on the two \ce{Mn} sublattices are presented in Fig.\ \ref{fig:induced_moments} as a function of the electric field direction. The orbital moments $\mu_L$ are always induced perpendicular to the electric field direction and are antisymmetric (staggered) for the \ce{Mn} atoms of two sublattices. The spin moments $\mu_S$, on the other hand, are not necessarily perpendicular to the electric field direction, but their in-plane components are staggered as well. Additionally, the spin moments display a homogeneous out-of-plane component, i.e., a non-N\'eel-type contribution. 

Interestingly, in all configurations the induced orbital moments are more than one order of magnitude larger than the induced spin moments, yet the former were not included in previous studies \cite{zelezny_relativistic_2014,zelezny_spin-orbit_2017}.
To summarize, there are always quite large staggered orbital moments induced on the \ce{Mn} sublattices and small induced spin moments with nonstaggered as well as staggered components that can be parallel on antiparallel to the orbital moments depending on the direction of the electric field, see also \cite{salemi_orbitally_2019}.

\section{Atomistic spin dynamics simulations} 

To include our first-principles calculations in a spin dynamics simulation we extend the semi-classical Heisenberg Hamiltonian [Eq.\ (\ref{eq:spin_model})] by contributions from induced spin and orbital moments,
\begin{equation}
\begin{aligned}
	\mathcal{H} =&
	- \frac{1}{2}\sum_{i\neq j}J_{ij}(\vec{S}_{i}+\vec{s}_{i})\cdot(\vec{S}_{j}+\vec{s}_{j})\\
	& - \sum_{i}J_{}^{\mathrm{sd}}\vec{S}_{i}\cdot\vec{s}_{i}
	+ \sum_{i} \xi_{}\vec{S}_{i}\cdot\vec{l}_{i}\\
	&-\sum_{i} d_{z}S_{i,z}^2 -\sum_{i} d_{xy}S_{i,x}^2S_{i,y}^2 \ ,
\end{aligned}
\end{equation}
where $\vec{S}_{i} =\vec{\mu}^{\mathrm{d}}_{S,i}/{\mu}^{\mathrm{d}}_{S}$ is the local magnetic moment of the $\mathrm{d}$-electrons, $\vec{s}_{i} =\vec{\mu}^{\mathrm{s}}_{S,i}/{\mu}^{\mathrm{d}}_{S}$ the induced magnetic moment from the conduction $\mathrm{s}$-electrons and $\vec{l}_{i} =\vec{\mu}_{L,i}/{\mu}^{\mathrm{d}}_{S}$ the induced orbital magnetic moment. 
All magnetic moments are normalized with respect to the local magnetic moment.
Thus, the Hamiltonian consists of five different contributions: the inter-atomic exchange with exchange constant $J_{ij}$, an additional intra-atomic $\mathrm{sd}$-exchange with exchange constant $J^{\mathrm{sd}}_{}$, a \ac{soc} term with strength $\xi_{}$, as well as second and fourth order anisotropy terms constituting the tetragonal anisotropy. 

As our classical spin model employs quantum mechanical and statistical averages of the spin and orbital moments, we also use a classical description of the \ac{soc}  replacing the spin and orbital momentum operators by their averages. Note that this effective model for the \ac{soc} was used by Bruno \cite{bruno_physical_1993} in order to provide a simple physical interpretation of magnetic anisotropy.   
In this model only the spin moments couple via the inter-atomic exchange interaction, which is in agreement with the conclusions from \cite{sapozhnik_experimental_2018}.

All the contributions from the induced moments can also be represented by a simple Zeeman-like term with a sublattice-specific effective field that represents the staggered field, which was used in previous phenomenological descriptions,
\begin{align}
	{\mu}^{\mathrm{d}}_{S}\vec{B}_{i}^{\mathrm{ind}} =  \sum_{j}J_{ij}\vec{s}_{j} + J_{}^{\mathrm{sd}}\vec{s}_{i} - \xi_{}\vec{l}_{i} \ .
\end{align}
For the intra-atomic exchange we estimate from the shift in the up and down s-states $J^{\mathrm{sd}}=\SI{50}{\milli\electronvolt}$. The \ac{soc} strength is calculated from the energy difference between the $\mathrm{d}_{3/2}$ and $\mathrm{d}_{5/2}$ resonances yielding $\xi=\SI{46}{\milli\electronvolt}$. 
Together with the exchange interactions derived in Sec.\ II and the induced moment calculated for an electrical field of $\SI{1e7}{\volt\per\meter}$, this yields staggered fields of about $\SI{76}{\milli\tesla}$.
Here, the contribution from the induced orbital moments dominates \cite{salemi_orbitally_2019}. It is about a factor of five larger than the contribution from the inter-atomic exchange and more than one order of magnitude larger than that of the intra-atomic exchange.
This explains also why the staggered fields calculated here are much larger than those estimated and predicted before \cite{roy_robust_2016,zelezny_spin-orbit_2017,meinert_electrical_2018} as the orbital contribution was previously not taken into account.

The time evolution of the localized \ce{Mn} moments stemming from the $\mathrm{d}$-electrons is described by the stochastic \ac{llg} equation
\begin{align}
\dot{\vec{S}}_{i}=-\frac{\gamma}{\left(1+\alpha^2\right){\mu}^{\mathrm{d}}_{S}} \vec{S}_{i}\times \Big[\vec{H}_{i}+\alpha \vec{S}_{i}\times\vec{H}_{i}\Big] \ ,
\end{align}
where $\gamma=\SI{1.76e11}{\per\second\tesla}$ is the gyromagnetic ratio and $\alpha$ a dimensionless damping constant. Temperature is included via Langevin dynamics by adding a random thermal noise $\vec{\zeta}_{i}$ to the effective field $\vec{H}_{i}=-\frac{\partial\mathcal{H}}{\partial\vec{S}_i}+\vec{\zeta}_{i}$ \cite{kronmuller_classical_2007}. The field from the induced moments $\vec{B}_{i}^{\mathrm{ind}}$ is part of this effective field.

The damping constant is a free parameter as there are no experimental values for it in the literature.
For comparison with \cite{roy_robust_2016} we use a plausible value of $\alpha=0.01$. Similar, for the electrical field a rectangular pulse with pulse length of \SI{20}{\pico\second} was simulated to compare the results with those from a phenomenological model \cite{roy_robust_2016}.
Since the samples in experiments are mostly of granular type \cite{meinert_electrical_2018}, we simulate a system of $\SI{20.3}{\nano\meter}\times\SI{20.3}{\nano\meter}\times\SI{20.5}{\nano\meter}$ size with open boundary conditions, resembling one grain of a typical sample.

\begin{figure}
	\centering
	\includegraphics[width=\linewidth]{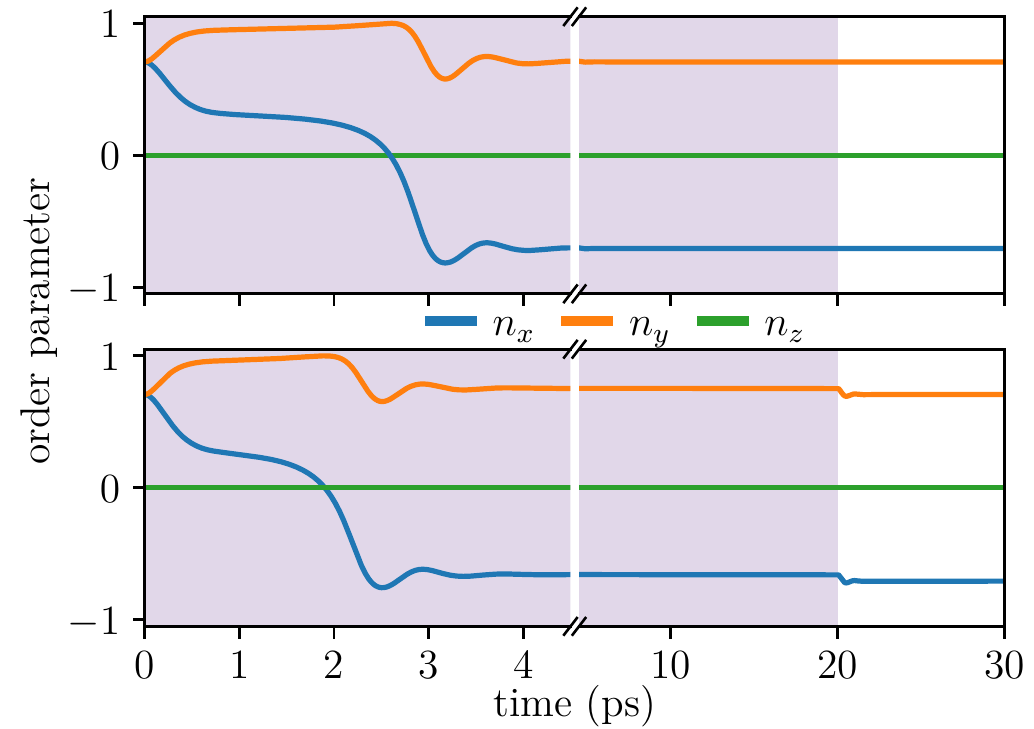}
	\caption{Time evolution of the magnetic order parameter during $90^{\circ}$ switching at $T=\SI{0}{\kelvin}$. The electric field (applied in the shaded area) is \SI{1.9e7}{\volt\per\meter} in \mbox{[110]} direction (top) and \SI{3.1e7}{\volt\per\meter} in \mbox{[100]} direction (bottom).\label{fig:T0_comparison}}
\end{figure}

In our simulations we consider electrical fields along \mbox{[110]}, i.e.\ parallel to the local magnetic moments, and along \mbox{[100]}, since reversible switching was reported for both directions \cite{bodnar_writing_2018}.
For both field configurations our model does not switch at $T=0$ for  $E=\SI{1e7}{\volt\per\meter}$ corresponding to currents of about \SIrange{1e10}{1e11}{\ampere\per\meter\squared}, which are used in experiments \cite{bodnar_writing_2018,bodnar_imaging_2019}. Instead, we need a field strength of at least  $E=\SI{1.9e7}{\volt\per\meter}$ for the field along the \mbox{[110]} direction, where torques on the local magnetic moments are maximal. For the \mbox{[100]} direction an even larger field of $\SI{3.1e7}{\volt\per\meter}$ is required for switching at zero temperature. However, once  the system switches, it switches within a few picoseconds, see Fig.~\ref{fig:T0_comparison}. This is even faster than predicted in the phenomenological model in Ref.~\cite{roy_robust_2016}, probably because of the inclusion of the orbital induced moments and the exchange interactions beyond the first three nearest neighbors.

\begin{figure}
	\centering
	\includegraphics[width=0.9\linewidth]{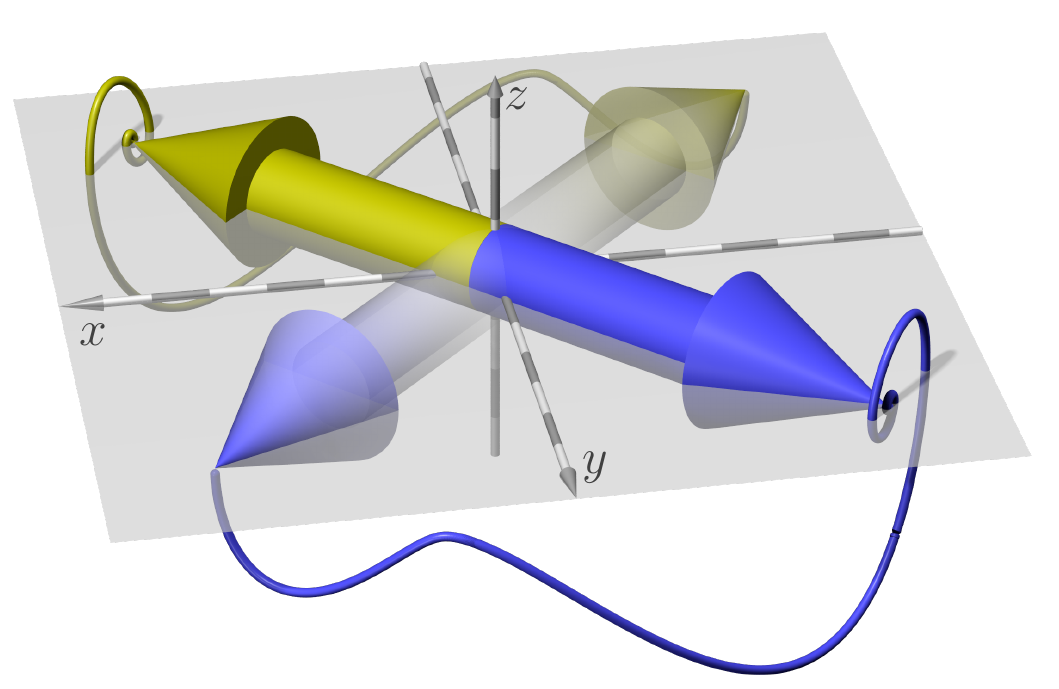}
	\caption{Switching path of the two sublattice magnetization vectors. The antiparallel \ce{Mn} moments switch over 90$^{\circ}$ from the initial \mbox{[110]} configuration (semi-transparent) to the final \mbox{[-110]} configuration (opaque). During the switching process the sublattices are canted slightly resulting in huge torques from the inter-sublattice exchange field enhancing the switching process significantly. This exchange enhancement is characteristic for antiferromagnetic dynamics \cite{dannegger2021}. The out-of-plane component is here scaled by a factor of 100!
	\label{fig:switching_path}}
\end{figure}

As was already pointed out in Ref.~\cite{roy_robust_2016}, the reason for this rapid switching is the so-called exchange enhancement, which is characteristic for antiferromagnetic dynamics \cite{dannegger2021}. The staggered fields do not only rotate the magnetic moments via the damping term in the \ac{llg} but also induce a canting between the sublattices via the much stronger precession term. 
This leads to a very small magnetization resulting in huge torques due to the inter-sublattice exchange field, which tries to realign the sublattices. Here, the damping term is responsible for the realignment. The precession term, on the other hand, rotates the magnetic moments towards the direction of the staggered field. The out-of-plane component of the order parameter remains zero during the process (see Fig.\ \ref{fig:switching_path}). Hence, the inter-sublattice exchange field governs the switching process and, in contrast to the switching in \acp{fm}, lower damping allows for faster switching  \cite{rozsa_reduced_2019}.

\begin{figure}
	\centering
	\includegraphics[width=\linewidth]{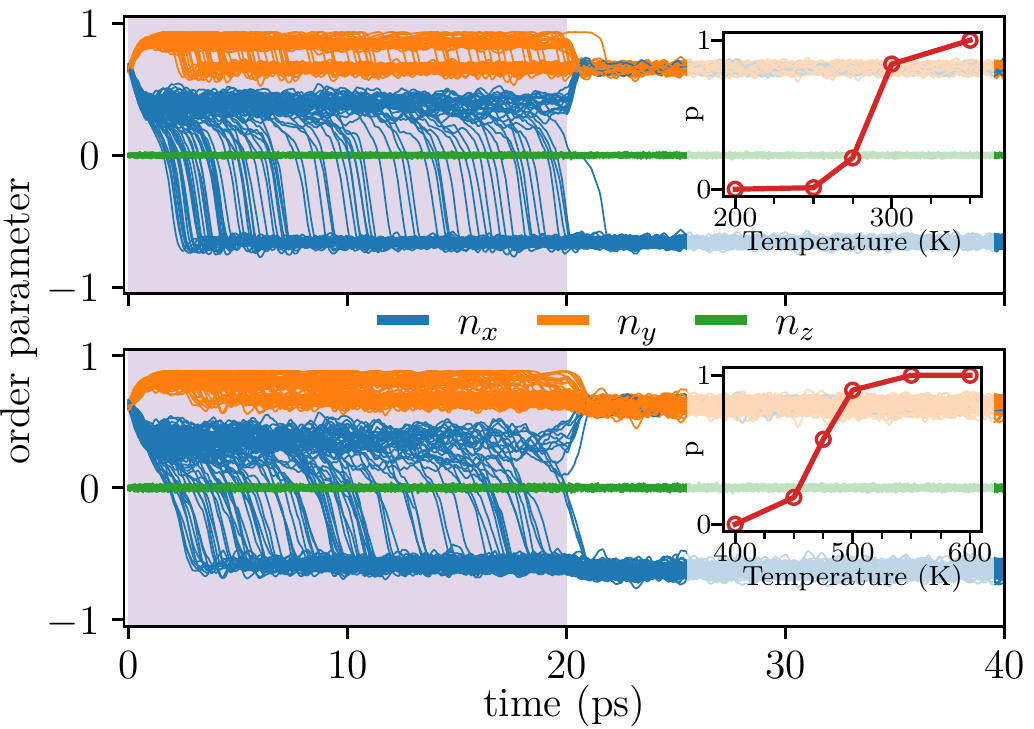}
	\caption{100 trajectories of the order parameter during the electric field pulse of $E=\SI{1e7}{\volt\per\meter}$ (shaded area). Top: at $T=\SI{300}{\kelvin}$ for an electric field in \mbox{[110]} direction. Bottom: at $T=\SI{500}{\kelvin}$ for an electric field in \mbox{[100]} direction. The inset shows the switching probability as a function of temperature.\label{fig:T_comparison}}
\end{figure}

The electric fields considered so far are much larger than those applied in experiments, but  temperature plays an additional major role. A finite temperature does not only lower the energy barrier, here the fourth order in-plane magnetic anisotropy, but thermal fluctuations can also support  probabilistic switching.
Fig.~\ref{fig:T_comparison} shows the time evolution of the order parameter at elevated temperatures as well as the switching probability as a function of temperature for electrical fields of $E=\SI{1e7}{\volt\per\meter}$. For the \mbox{[110]} direction the system does not switch at temperatures below \SI{250}{\kelvin}, between \SI{250}{\kelvin} and \SI{350}{\kelvin} the process is probabilistic and above \SI{350}{\kelvin} deterministic.
In the deterministic regime the energy barrier is so low that the system switches in a few picoseconds, similar to simulations with larger electric fields. In the probabilistic regime, however, it can take several attempts to cross the energy barrier due to thermal agitation. Of course, here the switching probability also depends on the pulse length of the external electric field as longer time scales allow for more stochastic attempts to cross the barrier.
For the electric field along the \mbox{[100]} direction the probabilistic regime lies between \SI{400}{\kelvin} and \SI{550}{\kelvin}, above which the switching is deterministic.

Reversible switching for pulse currents along the \mbox{[100]} direction was also observed in experiments \cite{bodnar_writing_2018}. In the same paper, also significant heating resulting in temperatures up to \SI{300}{\celsius} was reported and thermal activation was considered to play an important role in the process. A key role of thermal activation was also reported by \textcite{meinert_electrical_2018}.
Of course, for thermal switching of nanoparticles the system size is crucial as well, especially for antiferromagnets as their thermal stability is much lower than that of ferromagnets \cite{rozsa_reduced_2019}. Here, the system size was chosen such to avoid  a purely superparamagnetic switching which would lead to a forth and back switching.

\section{Conclusions} 

Modeling the current induced switching process in \ce{Mn2Au} with all its different contributing terms in a quantitative manner is a challenging task. Here, we have presented the first multi-scale model combining first-principles calculations of exchange and anisotropy constants, as well as electrically induced spin and orbital moments in an extended atomistic spin model.
We predict much higher effective staggered fields due to the formerly neglected contributions from induced orbital moments.
Within the framework of atomistic spin dynamics simulations, we have shown that these fields---combined with inter-sublattice exchange interactions---result in switching processes on the time scale of few picoseconds. However, this switching requires significantly higher electrical fields than in experiments or, alternatively, elevated temperatures. This applies for both considered electrical field directions, \mbox{[110]} and \mbox{[100]}, which is in agreement with experimental findings \cite{bodnar_writing_2018}.
Hence, in agreement with previous experimental studies \cite{bodnar_writing_2018,meinert_electrical_2018} we find that thermal activation plays a key role in the current induced switching process and we have consequently distinguished temperature regimes for probabilistic and deterministic switching.

\begin{acknowledgments}
The authors gratefully acknowledge valuable discussions with Karel Carva. L.S.\ and P.M.O.\ acknowledge funding from the Swedish Research Council (VR) and the European Union’s Horizon2020 Research and Innovation Programme under FET-OPEN Grant agreement No.\ 863155 (s-Nebula), and acknowledge computer resources provided by the Swedish National Infrastructure for Computing (SNIC) at the PDC Center for High Performance Computing and the Uppsala Multidisciplinary Center for Advanced Computational Science (UPPMAX). The work of A.D., E.S.\ and L.Sz.\ was supported by National Research, Development, and Innovation Office under projects
No.\ PD134579 and No.\ K131938. The work in Konstanz was supported by the {Deutsche Forschungsgemeinschaft} via the {Sonderforschungsbereich 1432}.
\end{acknowledgments}

\bibliography{literature}

\end{document}